\documentclass[aps,prb,twocolumn,groupedaddress,showpacs,floatfix,altaffilletter]{revtex4-1}
\usepackage{graphicx}
\usepackage{grffile}
\usepackage{amsmath,amsfonts}
\usepackage{amssymb}
\usepackage{bm}
\usepackage{color}
\usepackage[dvipsnames]{xcolor}
\usepackage{amsmath,amssymb,amsfonts}
\usepackage{epsfig}
\usepackage{times}
\usepackage[colorlinks,bookmarks=false,citecolor=blue,linkcolor=red,urlcolor=blue]{hyperref}

\newcommand{\fref}[1]{Fig.~\ref{#1}}
\newcommand{\eref}[1]{Eq.~(\ref{#1})}

\newcommand{\normwidth}{0.8\columnwidth}

\newcommand{\figwidth}{0.97\columnwidth}

\begin{document}

\title{Orbital magnetization of correlated electrons with arbitrary band topology}

\author{R.~Nourafkan$^{1,2}$, G.~Kotliar$^{1}$, and A.-M.S. Tremblay$^{2,3}$}
\affiliation{$^{1}$Department of Physics \& Astronomy, Rutgers University, Piscataway, NJ 08854-8019, USA}
\affiliation{$^2$D{\'e}partement de Physique and RQMP, Universit{\'e} de Sherbrooke, Sherbrooke, Qu{\'e}bec, Canada}
\affiliation{$^3$Canadian Institute for Advanced Research, Toronto, Ontario, Canada}
\date{\today}
\begin{abstract}
Spin-orbit coupling introduces chirality into the electronic structure. This can have profound effects on the magnetization induced by orbital motion of electrons. Here we derive a formula for the orbital magnetization of interacting electrons in terms of the full Green function and vertex functions. The formula is applied within dynamical mean-field theory to the Kane-Mele-Hubbard model that allows both topological and trivial insulating phases. We study the insulating and metallic phases in the presence of an exchange magnetic field. In the presence of interactions, the orbital magnetization of the quantum spin Hall insulating phase with inversion symmetry is renormalized by the bulk quasi-particle weight. The orbital magnetization vanishes for the in-plane antiferromagnetic phase with trivial topology. In the metallic phase, the enhanced effective spin-orbit coupling due to the interaction sometimes leads to an enhancement of the orbital magnetization. However, at low doping, magnetization is suppressed at large interaction strengths.
\end{abstract}
\pacs{75.10.Lp, 71.10.Fd, 03.65.Vf, 75.20.-g}
%75.00.00 Magnetic properties and materials
%75.10.Lp Band and itinerant models
%71.27.+a Strongly correlated electron systems; heavy fermions
%03.65.Vf Phases: geometric; dynamic or topological
%71.30.+h Metal-insulator transitions and other electronic transitions
%71.10.Fd Lattice fermion models (Hubbard model, etc.)
%75.20.-g Diamagnetism, paramagnetism, and superparamagnetism

\maketitle
\section{Introduction}
Magnetism of matter in thermal equilibrium is a purely quantum mechanical phenomenon. For conventional metals one usually identifies two contributions: a paramagnetic one -Pauli- due to the magnetic moment of the spin, and a diamagnetic one -Landau- due to the orbital motion of electrons.~\cite{Nolting+Ramakanth} In the free electron case, the magnitude of the spin contribution is larger by a factor of three compared with the orbital contribution so that the system exhibits paramagnetism. However, orbital magnetism depends sensitively on details of the electronic structure, and sometimes deviates strongly from conventional Landau diamagnetism. For instance, narrow gap materials such as bismuth~\cite{Nolting+Ramakanth} and graphene~\cite{RevModPhys.81.109} exhibit considerably enhanced diamagnetism. Also, the chirality imposed on electronic states by spin-orbit coupling leads to the appearance of new states of matter with peculiar magnetic responses. For example, it has recently been shown experimentally that some iridate compounds~\cite{PhysRevLett.105.216407, PhysRevB.85.224422} possess strong orbital magnetism that dominates over spin paramagnetism. 

The modern theory of orbital magnetization \cite{PhysRevLett.95.137204, PhysRevLett.95.137205, PhysRevB.74.024408, PhysRevLett.99.197202, PhysRevB.84.205137, PhysRevLett.110.087202} focuses on a crystalline system of independent electrons in a state that breaks time-reversal symmetry. In this theory, the orbital magnetization comes from the orbital motion of carriers and also from a correction due to the Berry curvature. It has become important to generalize this modern theory of orbital magnetization to include the effect of interactions. Indeed, the first principles application of the theory for ferromagnetic transition metals have shown that this theory underestimates the orbital magnetization.~\cite{PhysRevB.81.060409, PhysRevB.85.014435} It is reasonable to expect that interactions can explain this discrepancy. Also, interactions renormalize the electronic structure of the system, in some cases so drastically that they cause a phase transition. Interactions can therefore modify both contributions to the orbital  magnetization.

Here we derive a formula for the orbital magnetization of an interacting system in terms of the fully interacting Green function and of the corresponding vertex functions. 
The proposed formalism can be used for systems with arbitrary band topology along with any method capable of calculating the interacting Green function, such as \emph{GW} or DMFT. As a simple example, we apply this formula to the Kane-Mele-Hubbard (KMH) model~\cite{PhysRevB.85.115132} in the presence of an exchange magnetic field that acts on spins only to break time-reversal symmetry. We allow for a chiral symmetry breaking perturbation in the KMH so that we can study both the correlated topological insulating phase and the trivial insulating phase.

\section{Derivation}
The thermodynamic definition of the orbital magnetization density at zero temperature is,
\begin{equation}
{\bf M}_{orb}=-\left(\frac{\partial K}{\partial {\bf B}}\right)_{n,{\bf B}={\bm 0}},
\label{eq:morb1}
\end{equation}
where $K$ is the grand potential per unit volume of the system, $\bf B$ is a magnetic field and the derivative is evaluated at constant electron density. To focus on the orbital contribution, we exclude the Zeeman energy. The full algebraic derivation is given in appendix A. But it is in fact simple to understand the procedure and final result. One cannot take directly the derivative with respect to a uniform magnetic field since, fundamentally, $K$ is a function of a vector potential that must depend on position (see also~\cite{PhysRevB.84.205137}). Hence, going to Fourier space, one must expand $K$ in powers of $q_b$ and $A_c$ and keep the part of the derivative that is antisymmetric under exchange of the cartesian directions $b$ and $c$. Computing $\epsilon^{abc}\frac{\partial^2K}{\partial i q_b \partial A_c}(\epsilon^{ade} i q_dA_e)$ with $\epsilon^{abc}$ the fully antisymmetric Levi-Civita tensor, we thus obtain $2\frac{\partial K}{\partial B_a}B_a$. 
The expression for $K$ in presence of the gauge field ${\bf A}({\bf q})$ involves an energy vertex multiplied by a dressed Green function that depends on two wave vector indices, ${\bf k}-{\bf q}/2$ and ${\bf k}+{\bf q}/2$, since we do not have translational invariance (see \eref{eq:K00app} in appendix A).  That Green function depends implicitly on ${\bf A}$, which also appears in the energy vertex through the usual Peierls substitution. Taking derivatives with respect to $q_b$ and $A_c$ and taking the anti-symmetric part in the limit of zero field, we obtain the orbital magnetization. 
%The procedure is illustrated in \fref{feyndiag} in terms of dressed Feynman diagrams. 
Algebraically, one finds
\begin{widetext}
\begin{eqnarray}
M^a_{orb}&=&(\frac{ie}{2\hbar})(\frac{1}{N\beta})\sum_{{\bf k},\omega_m} \epsilon^{abc}{\rm Tr}\bigg([{\bf H}_0-\mu{\bm 1}+\frac{{\bm \Sigma}}{2}]{\bf G}(-\frac{\partial {\bf G}^{-1}}{\partial k_b}) {\bf G}(-\frac{\partial {\bf G}^{-1}}{\partial k_c}){\bf G}\bigg)e^{i\omega_m 0^+}\nonumber\\&+&
(\frac{1}{2N\beta})\sum_{{\bf k},\omega_m} {\rm Tr}\bigg([{\bf H}_0+(i\omega_m-\mu){\bm 1}]{\bf G}\left(\frac{\partial {\bm \Sigma}^{({\bm B})}}{\partial B_a}\right)_{{\bf B}={\bm 0}} {\bf G}\bigg).
\label{eq:morb3}
\end{eqnarray}
\end{widetext}
Derivatives with respect to $k_i$ appear because in the zero-field limit, derivatives with respect to $A_i$ or to $q_i$ are proportional to $\frac{\partial}{\partial k_i}$. The identity $\frac{\partial \mathbf{G}}{\partial k_b}=-\mathbf{G}\frac{\partial \mathbf{G^{-1}}}{\partial k_b}\mathbf{G}$ has been used repeatedly. The interacting single-particle Greens function entering \eref{eq:morb3} is
\begin{equation}
{\bf G}({\bf k},i\omega_m)=[(i\omega_m+\mu){\bm 1}-{\bf H}_0({\bf k})-{\bm \Sigma}({\bf k},i\omega_m)]^{-1}, 
\end{equation}
where ${\bf H}_0$ denotes the non-interacting part of Hamiltonian, ${\bm \Sigma}$ is the electron self-energy, $\beta$ is the inverse temperature, $\mu$ is the chemical potential and $\omega_m$ denotes the Matsubara frequencies. 
Bold quantities are written in spinor notation and their size is $2n\times 2n$ where $n$ denotes the number of orbitals within the unit cell. 

Equation (\ref{eq:morb3}) is an antisymmetric response that cannot be attributed to Lorentz forces and therefore survives in the absence of a magnetic field. It is valid for both trivial and topological insulators as well as for metals.  In the noninteracting case \eref{eq:morb3} reduces to the modern theory of  orbital magnetization (see appendix B). We apply \eref{eq:morb3} then to the KMH model with a chiral symmetry breaking term as an example that will illustrate the effect of interactions.

\section{Kane-Mele-Hubbard model}
The Hamiltonian on the honeycomb lattice reads
\begin{eqnarray}
H&=&-t\sum_{\langle ij\rangle} \hat{c}^{\dagger}_{i}{\bm 1}\hat{c}_{j}+i\lambda_{SO}\sum_{\langle\langle ij \rangle\rangle }\hat{c}^{\dagger}_i{\bm \tau}\cdot({\bm \delta}^{(1)}_{ij}\times{\bm \delta}^{(2)}_{ij}) \hat{c}_j\nonumber\\&-&\lambda(\sum_{i\in A}\hat{c}^{\dagger}_i {\bm 1} \hat{c}_i-\sum_{i\in B}\hat{c}^{\dagger}_i {\bm 1} \hat{c}_i)+\frac{U}{2}\sum_i(\hat{c}^{\dagger}_i{\bm 1}\hat{c}_i-1)^2,\label{eq:KMH}
\end{eqnarray}
where $\hat{c}^{\dagger}_{i}\equiv({c^{\dagger}_{i\uparrow},c^{\dagger}_{i\downarrow}})$ is a spinor and $c^{\dagger}_{i\uparrow}$ creates an electron with spin $\sigma$ on site $i$.  The second term is a mirror symmetric ($z\rightarrow -z$) spin-orbit interaction, which involves  spin-dependent hopping between pairs of second neighbors $\langle\langle ij \rangle\rangle$,  with ${\bm \delta}^{(1,2)}_{ij}$ the vectors connecting first-neighbor legs and $\bm \tau$ the Pauli spin matrices.~\cite{PhysRevB.85.115132} 

We use Dynamical Mean Field Theory (DMFT) with two single-site impurity models per unit cell.~\cite{RevModPhys.68.13} Thus the self-energy is a block-diagonal matrix with $2\times 2$ elements ${\bm \Sigma}_{A}, {\bm \Sigma}_{B}$ in spin-space. We use an exact diagonalization impurity solver~\cite{PhysRevLett.72.1545} with 8 bath sites. To treat long-range in-plane antiferromagnetic order, we add a self-consistent Weiss field to the bath.~\cite{PhysRevLett.107.010401} As a check of the accuracy of the method, we compare our DMFT calculation with those obtained from a quantum Monte Carlo study.~\cite{PhysRevB.85.115132} We find that the critical values of $U_c$ for the transition between the QSH and the antiferromagnetic (AFM) phase are within a few percent of each other, and similarly for the value of the single-particle gap for $\lambda_{SO}=0.1t$. 

In the DMFT approximation, the current vertex corrections from $\partial {\bf \Sigma}/\partial k_{b,c}$ vanish and since the scalar ${\bm \Sigma}^{({\bf B})}$ is independent of ${\bf k}$, it cannot depend on ${\bf B}$ linearly so $\partial {\bm \Sigma}^{({\bf B})}/\partial B_a=0$. 

At half-filling, the noninteracting system with inversion symmetry  ($\lambda_{SO}\ne 0, \lambda=0$) describes a Quantum Spin Hall (QSH) insulator with helical edge states.
In the system without inversion symmetry, ($\lambda \ne 0$), a phase transition between the QSH insulator and a band insulator occurs at $\lambda > 3\sqrt{3}\lambda_{SO}$.\cite{PhysRevB.85.115132} 

The Hubbard repulsion
induces a transition from the  correlated QSH phase to a
Mott insulator with long-range in-plane antiferromagnetic order at a critical
value. 
~\cite{PhysRevB.85.115132, PhysRevB.85.205102} (see appendix C) Throughout the QSH phase, the bulk gap remains open. At the magnetic transition, the time-reversal symmetry underlying the topological protection of the QSH state is broken: A change of the topological invariant from nontrivial to trivial does not require the closing of any gaps.~\cite{PhysRevB.85.115132}  

In the correlated QSH insulator, time-reversal symmetry is preserved and therefore the net orbital magnetization is zero. Nevertheless, the integrand ${\bf m}_{orb}({\bf k})$ in the general result \eref{eq:morb3} has a strong ${\bf k}$ and $\mu$ dependence.  We first study its behaviour in the noninteracting case since it contains many features that
remain in the interacting system.  

\section{Results} 
\subsection{Noninteracting case:} 
In the noninteracting system with a chemical potential in the band gap, one can use the low-energy description near the Dirac points to obtain an approximate analytical expression for the Berry curvature correction contribution (see appendix D),
\begin{eqnarray}
m^{Berry}_{orb}({\bf q})&=&(\frac{e}{4\hbar })\sum_{s,s_v}[(\Delta^{s2}_v+\hbar^2 v_F^2q^2)^{1/2}+\mu] \nonumber \\
&\times& \frac{s^v\Delta^s_v\hbar^2 v_F^2}{[\Delta^{s2}_v+\hbar^2 v_F^2q^2]^{3/2}}.\label{eq:analm}
\end{eqnarray}
where 
${m}^{Berry}_{orb}({\bf q})$ is the magnitude of ${\bf m}^{Berry}_{orb}({\bf q})$,
$\Delta^s_v=(-\lambda+ss^v3\sqrt{3}\lambda_{SO})$ is a valley and spin dependent gap, $\hbar v_F=(3/2)at$ is the Fermi velocity of the helical Dirac fermions, $\bf q$ is in the neighbourhood of the valley, $s^v=\pm 1 $ is the pseudospin valley index, and $s=\pm 1$ is the electron spin index. The orbital moment contribution has similar structure. 

Consider first a trivial insulator, $\lambda>3\sqrt{3}\lambda_{SO}$. Since $\Delta^s_v$ has a valley and spin independent sign, Eq. (\ref{eq:analm}) shows that the orbital magnetization integrand within each band has opposite sign in the two valleys.~\cite{PhysRevLett.99.236809} Even though each band has states with both chirality, in the presence of a non-zero $\lambda_{SO}$, states with opposite chirality are not balanced and each band has a net chirality. When $\lambda_{SO}$ vanishes, states with opposite chirality balance each other and the net orbital magnetization of each band is individually zero: A response of type \eref{eq:morb3} is not present in this case. 

For the topological insulator (QSH) with inversion symmetry, $\lambda_{SO}\ne 0,\lambda=0$, the $m^{Berry}_{orb}({\bf k})$ of each band has the same sign for the two valleys, i.e., for a given spin, each band has only states with a specific chirality, giving rise to a large contribution to orbital magnetization. A small $\lambda$ breaks the symmetry between the two valleys. 

For both band insulator and QSH insulator, $m^{Berry}_{orb}({\bf q})$ of the valence (conduction) bands (summed over spin) have opposite sign as required by the fact that the KM model preserves time-reversal symmetry and therefore the orbital magnetization is zero.   

\begin{figure}
\begin{center}
\includegraphics[width=\figwidth]{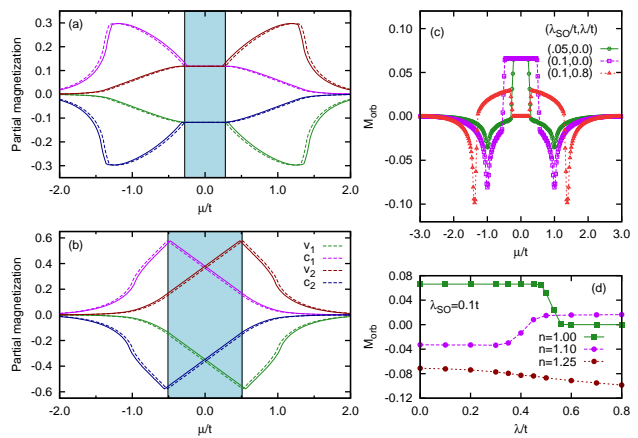}
\caption{(color online) Orbital magnetization for the non-interacting KM model. Panels (a) and (b) show the partial (band) orbital magnetization for, respectively, the trivial band insulating phase, $\lambda_{SO}=0.1t, \lambda=0.8t$, and the QSH insulating phase, $\lambda_{SO}=0.1t, \lambda=0$. The partial orbital magnetization in the presence of a time-reversal symmetry-breaking exchange field $h=-0.04t$ acting on spins only are shown by solid lines. Dashed lines show the partial orbital magnetization in absence of an exchange field. The shaded area shows the bulk spectrum gap. Symbols for valence and conduction bands are identified in panel (b). In panels (c) and (d), $h=-0.04t$. The total orbital magnetization as a function of $\mu$ is in (c). In (d) orbital magnetization with $\lambda_{SO}=0.1t$ as a function of $\lambda/t$ for electron densities $n=1.0, 1.1$ and $n=1.25$. The semi-metal phase at the boundary between QSH insulator and the trivial band insulator is broadened slightly by the applied exchange field. All data are in units of $(ea^2t/2\hbar)$ where $a$ is the lattice constant.}\label{fig6}
\end{center}
\end{figure}

A numerical evaluation of \eref{eq:morb3} with the full Green functions confirms the above analysis based on the Dirac approximation: Panels (a) and (b) of \fref{fig6} show the partial orbital magnetization contribution of each band in the trace entering \eref{eq:morb3} as a function of chemical potential. In the band insulator, \fref{fig6}(a),  the partial orbital magnetization is constant for a chemical potential lying in the gap (shaded area) while it linearly changes in the QSH insulator , \fref{fig6}(b), with a slope proportional to the Chern number of the band. \cite{PhysRevB.74.024408} This can be interpreted as an effect due to populating the edge states. Although there is no edge in an extended system, this demonstrates that the bulk response can be encoded in the boundary, as expected from  bulk-boundary duality. \cite{RevModPhys.82.3045} In the band insulator the absolute value of the partial orbital magnetization of each band increases when $\mu$ increases outside the gap, reaches a maximum once $\mu$ is at the energy of the van Hove singularity of the corresponding band and then decreases for larger chemical potentials.       

Applying a small uniform exchange (Zeeman) field, $-h\sum_{i}\hat{c}^{\dagger}_i {\bm \tau}_z \hat{c}_i$, perpendicular to the plane, breaks time-reversal symmetry and mirror symmetry with respect to the plane and induces a non-zero orbital magnetization.~\footnote{Here we ignore the direct influence of the magnetic field on the phase of the itinerant electrons coming from conventional diamagnetism.} At small field strengths, the QSH state survives, regardless of the broken time-reversal symmetry.~\cite{PhysRevLett.107.066602} The variation in the orbital magnetization is given by the difference between the Bloch states carrying circulating currents in opposite directions. A non-zero $h$ shifts the energy of the Kramer's pair bands relative to each other and creates these differences. Figure \ref{fig6}, panels (a) and (b) show how the exchange field breaks the balance between Bloch states carrying opposite circulating currents.  

Panel (c) of \fref{fig6} shows the orbital magnetization of the KM model in the topological and trivial phases in presence of an exchange field. The direction of the orbital magnetization depends on the sign of $\lambda_{SO}$ and of $h$. As can be seen from the figure (green and purple lines) in the topological insulator the orbital magnetization is independent of Hamiltonian parameters. This can be understood as follows: In the insulating phase only the Berry curvature correction contributes to the net orbital magnetization. The applied Zeeman term does not change the Berry curvature of the bands, ${\Omega}_s({\bf q})$. However it linearly changes the energy vertex, $-sh$, in the Berry curvature correction of the orbital magnetization. Thus the net orbital magnetization due to the field is $h\sum_{{\bf q},s}{\Omega}_s({\bf q})$. The orbital magnetization is also independent from the position of the chemical potential in the gap. Scanning $\mu$ in the gap does not cause any change in the orbital magnetization due to presence of opposite Chern indices in the QSH insulator.  

The orbital magnetization of the trivial insulator (red line in \fref{fig6}(c)) is zero, meaning that for each Bloch state there is another state carrying opposite-circulating current. However, note that a trivial insulator with vanishing Chern index can in general have a small but finite orbital magnetization. Indeed, in the non-interacting case the energy vertex in \eref{eq:morb3} makes the expression for orbital magnetization different from that for the Chern index. In the trivial insulator phase of the KMH the following two conditions make the orbital magnetization vanish: particle-hole symmetry and $\mathbf{k}$-independence of the correction to the energy vertex due to the exchange field. 

Away from half-filing the orbital magnetization shows a complex structure that arises from both contributions of the  orbital magnetization. Nevertheless, the behaviour can be understood by inspecting \fref{fig6}(a) and (b). Comparing green and purple lines in \fref{fig6}(c) shows that in the metallic phase of the doped topological insulator, the absolute value of the orbital magnetization takes larger values upon increasing the spin-orbit coupling. 

Finally, \fref{fig6}(d) shows $M_{orb}$ as a function of staggered ionic potential, $\lambda/t$, for electron densities $n=1.0,1.1$ and $n=1.25$. 
At small doping level, $n=1.1$, the response changes from paramagnetic (diamagnetic) to diamagnetic (paramagnetic) as $\lambda/t$ increases, reflecting the crossover from a doped QSH to a doped band insulator. At higher doping level, $n=1.25$, only the magnitude  of the response changes when $\lambda/t$ increases.

\subsection{Interacting case:} In general, electronic correlations enhance the effects of spin-orbit coupling, due to the suppression of the effective bandwidth.~\cite{NaturePhys.6.376} This can be seen in a system with staggered sublattice potential where the real part of the self-energy renormalizes $\lambda \rightarrow \lambda^{ren} <\lambda$, increasing the stability of the topological insulator with increasing interaction. 

\begin{figure}
\begin{center}
\includegraphics[width=\figwidth]{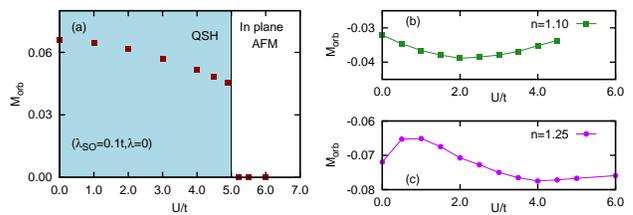}
\caption{(color online) $M_{orb}$ of the interacting KMH model as a function of $U/t$. Panel (a) at half-filling. The shaded area shows the correlated QSH phase. In panel (b), $M_{orb}$ with $\lambda_{SO}=0.1t, \lambda=0$ as a function of $U/t$ for electron densities $n=1.1$ (top) and $n=1.25$ (bottom). A small exchange field, $h=-0.04t$, is applied. 
There is an out-plane AFM phase for $n=1.1$ at $U/t \simeq 5.4$.}\label{fig8}
\end{center}
\end{figure}

\fref{fig8}(a) shows the orbital magnetization of the correlated QSH insulator ($\lambda_{SO}=0.1t,\lambda=0$) in the presence of a small exchange field $h$, as a function of interaction strength $U$. The interaction suppresses the orbital magnetization. This can be explained as follows. 
Although the time-reversal symmetry forbids elastic single-particle scattering  processes, two-particle scattering renormalizes the velocity.~\cite{PhysRevB.73.045322, PhysRevLett.107.010401} Within DMFT, one finds $v_F^{ren} \simeq zv_F$, where $z$ is the quasiparticle weight. The small exchange field does not change the scattering processes very much and this renormalization is valid even in presence of the field. Also, the band gap smoothly evolves from its $U/t=0$ value to its renormalized value $\Delta_{v}^{s,ren}=z[\Delta_{v}^s-\Re(\Sigma_{A,s}(0)-\Sigma_{B,s}(0))]$~\cite{PhysRevB.88.155121} With inversion symmetry, the zero-frequency self-energies cancel and we have $\Delta_{v}^{s,ren}\simeq z\Delta_{v}^s$.  
We can then use the quasi-particle Hamiltonian, ${\bf H}^{qp}_0={\bf z}^{1/2}({\bf H}_0-\Re {\bm \Sigma}(0)-\mu{\bm 1}){\bf z}^{1/2}$, with $\bf z$ the diagonal matrix of bulk quasi-particle weights with $z_A=z_B\equiv z$, to describe the correlated QSH insulator. Then, the Berry curvature of the correlated QSH insulator is given by second line of the \eref{eq:analm}, except that the bare quantities are replaced by renormalized ones, $\Delta_{v}^s \rightarrow \Delta_{v}^{s,ren}$ and $v_F\rightarrow v_F^{ren}$. Replacing the renormalized quantities in the Berry curvature equation, one can see that the quasi-particle weight cancel out from the equation and one find unrenormalized Berry curvature for the interacting case. However, in presence of the  interaction, the energy vertex renomalized as well. This renormalization leads to a suppression of the orbital magnetization. Using the numerically obtained value of $z$, we verified that the orbital magnetization is renormalized by the quasi-particle weight.   

Like the spin component, the net $M_{orb}$ would be zero for any AFM phase. Furthermore, although the orbital magnetization integrand may change drastically in the xy-AFM phase of KMH model, even in the presence of the exchange field the orbital magnetization vanishes because it is a trivial insulator. A study of AFM-Mott insulating perovskite transition metal oxides with a small net ferromagnetic (FM) moment using the modern theory of orbital magnetization have shown similar results.~\cite{PhysRevB.89.064428}

The right-hand panel of Fig. (\ref{fig8}) shows $M_{orb}$ of the interacting doped QSH with $\lambda_{SO}=0.1t, \lambda=0$ at $h\ne 0$ as a function of $U/t$. The early drop with $U/t$ of $|M_{orb}|$ at $n=1.25$ is due to the shift of the Fermi energy with respect to the rounded van Hove singularity. It does not occur at $n=1.1$. Then, as a function of $U/t$ the effective enhancement of $\lambda_{SO}$ leads to an increase in $|M_{orb}|$ but, eventually, at large $U/t$ the interaction effects described in the insulator lead to a net decrease in $|M_{orb}|$. 

\section{Conclusions:}
In conclusion, we have introduced a practical many-body approach for the
calculation of the orbital magnetization $|M_{orb}|$ of interacting systems with chiral electronic states. Using the Kane-Mele-Hubbard model in the presence of an exchange field as an example, we have shown that in the correlated topological insulator, $|M_{orb}|$ is decreased by the bulk quasi-particle weight $z$. In the doped topological insulator, the behavior of $|M_{orb}|$ is non-monotonic. Interaction effectively enhances the spin-orbit coupling and in turn the orbital magnetization while at the same time introducing scattering processes which reduce the orbital magnetization. Interplay between these two mechanism determine the orbital magnetization of a correlated system. The proposed formalism can be used for real material calculations along with any method capable of calculating the interacting Green function.

\begin{acknowledgments}
We are grateful to I. Garate for useful discussions and critical reading of the manuscript. We are also indebted L.-F. Arsenault. This work has been supported by NSF DMR-1308141, by the Natural Sciences and Engineering Research Council of Canada (NSERC), and by the Tier I Canada Research Chair Program (A.-M.S.T.). Simulations were performed on computers provided by CFI, MELS, Calcul Qu\'ebec and Compute Canada.

\end{acknowledgments}

\appendix
\section{Orbital magnetization}\label{App:AppendixA}
Here we present two derivations for the formula that gives the orbital magnetization of an interacting system. The first one follows the presentation in the main text. The second one generalizes the method introduced in Ref.~\onlinecite{PhysRevB.84.205137} to interacting systems. The latter method is more compact but perhaps less intuitive. 

\subsection{Derivation I}
\begin{figure*}
\begin{center}
\includegraphics[width=1.3\columnwidth]{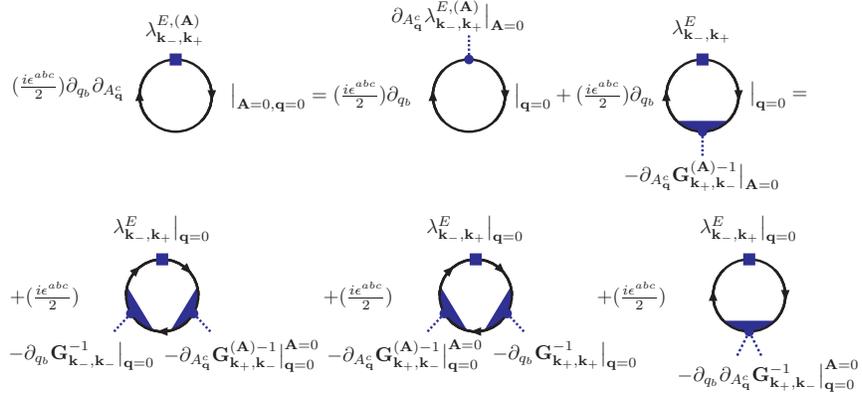}
\caption{(Color online) Diagrammatic expansion of the change in total energy due to the presence of a magnetic field, evaluated in the zero field limit. Lines show the fully dressed Green function, ${\bm \lambda}^{E}\equiv[{\bf H}_0+(i\omega_m-\mu){\bm 1}]$ is the energy vertex and ${\bf k}_-\equiv {\bf k}-{\bf q}/2$, ${\bf k}_+\equiv {\bf k}+{\bf q}/2$. The second diagram on the first line is independent of $\bf q$ and its derivative with respect to $\bf q$ vanishes. Evaluating the diagrams in the limit ${\bf q}\rightarrow {\bm 0}$ and ${\bf A}_{\bf q}\rightarrow {\bm 0}$ the derivative with respect to ${\bf A}_{\bf q}$ is replaced by $-(e/\hbar)\partial_{\bf k}$ while the derivative with respect to ${\bf q}$ is replaced by $(\pm 1/2)\partial_{\bf k}$ depending on the momentum of the propagator line. Two first diagram at the second line are equal in this limit and give the first line of \eref{eq:morb3}.}\label{feyndiag}
\end{center}
\end{figure*}

In this subsection, we provide details of the derivation for the orbital magnetization formula presented in the main text.  Since at the Hamiltonian level the magnetic field comes in through a vector potential ${\bf A}$, we must assume a long-wavelength variation of ${\bf A}({\bf r})={\bf A}_0\exp(i{\bf q}\cdot {\bf r})$, and take the limit ${\bf q}\rightarrow{\bm 0}$ at the end to recover a uniform magnetic field ${\bf B}$. The procedure is illustrated in \fref{feyndiag} in terms of dressed Feynman diagrams.

The variation of the total energy due to a small change in vector potential is
\begin{equation} 
\delta K\simeq\left(\frac{\partial K}{\partial {\bf A}_{\bf q}}\right)_{{\bf A}_{\bf q}={\bm 0}}\cdot \delta{\bf A}_{\bf q}.
\end{equation}
One then expands $\frac{\partial K}{\partial {\bf A}_{\bf q}}$ in powers of ${\bf q}$ up to linear order. Since ${\bf A}_{\bf q}$ itself is not expanded in powers of ${\bf q}$, one obtains:
\begin{equation}
\frac{\partial K}{\partial {\bf A}_{\bf q}}({\bf q}) \simeq 
\frac{\partial K}{\partial {\bf A}_{\bf q}}({\bf q}={\bm 0})+\mathbb{J}({\bf q}={\bm 0}){\bf q}+\cdots,\label{eq:Tylorapp}
\end{equation}
where $\mathbb{J}({\bf q}={\bm 0})$ is Jacobian matrix with the elements $\mathbb{J}_{bc}({\bf q}={\bm 0})=(\partial^2K/\partial q_b\partial A_c)_{{\bf q}={\bm 0}}$. The first term on the right-hand side of \eref{eq:Tylorapp} is zero because a uniform vector potential does not change the total energy of the system. Therefore, the first non-zero term in $\delta K$ in the limit of ${\bf q}\rightarrow{\bm 0}$ is 
\begin{eqnarray}
\delta K&=&\frac{1}{2}[ d{\bf A}_{\bf q}\cdot(\mathbb{J}({\bf q}={\bm 0}){\bf q})-d{\bf A}_{\bf q}\cdot(\mathbb{J}^{T}({\bf q}={\bm 0}){\bf q}) ]\nonumber\\&=&\frac{1}{2}\left({\bm\nabla}_{\bf q}\times{\frac{\partial K}{\partial {\bf A}_{\bf q}}}\right)_{{\bf q}={\bm 0}}\cdot  ({\bf q}\times d{\bf A}_{\bf q})\nonumber \\&=&\frac{i}{2}\left({\bm\nabla}_{\bf q}\times{\frac{\partial K}{\partial {\bf A}_{\bf q}}}\right)_{{\bf q}={\bm 0}}\cdot d{\bf B},\label{eq:dk2app}
\end{eqnarray}
where we have taken the anti-symmetric part on the right-hand side because the symmetric part contains contributions from pure gauge transformations, hence it cannot change the total energy. In the last identity we used the definition of the static magnetic field in terms of the vector potential, ${\bf B}({\bf q})=-i{\bf q}\times {\bf A}_{\bf q}$. 
Thus the orbital magnetization is given by,
\begin{eqnarray}
{\bf M}_{orb}&=&-\frac{i}{2}\left({\bm\nabla}_{\bf q}\times{\frac{\partial K}{\partial {\bf A}_{\bf q}}}\right)_{({\bf q},{\bf A})={\bm 0}}\label{eq:morb2app}
\end{eqnarray}
where it is understood that derivatives with respect to ${\bf q}$ do not act on ${\bf A}_{\bf q}$.

If we can compute the interacting Green's function ${\bf G}^{({\bf A})}$ in the presence of the space varying vector potential, the total energy per unit volume of the system can be calculated from
\begin{multline}
K=\frac{1}{2N\beta}\sum_{{\bf k}\omega_m} {\rm Tr}\bigg([{\bf H}^{({\bf A})}_{0,{\bf k}-{\bf q}/2,{\bf k}+{\bf q}/2}+(i\omega_m-\mu){\bm \delta}_{{\bf q},{\bm 0}}]\\{\bf G}^{({\bf A})}_{{\bf k}+{\bf q}/2,{\bf k}-{\bf q}/2}\bigg)e^{i\omega_m0^+},\label{eq:K00app}
\end{multline}    
where ${\bf H}^{({\bf A})}_0$ denotes the non-interacting part of Hamiltonian. It contains the vector potential through minimal coupling or through the Peierls substitution. The superscript $({\bf A})$ indicates that the quantity must be calculated in the presence of the field. The superscript is absent for quantities calculated at $\mathbf{B}=0$. In the presence of the non-uniform vector potential, the interacting Green's function ${\bf G}^{({\bf A})}$ depends on two wave-vectors. It takes the form
\begin{eqnarray}
{\bf G}^{({\bf A})}_{{\bf k}+{\bf q}/2,{\bf k}-{\bf q}/2}(i\omega_m)&=&[(i\omega_m+\mu){\bm \delta}_{{\bf q},{\bm 0}}-{\bf H}^{({\bf A})}_{0,{\bf k}+{\bf q}/2,{\bf k}-{\bf q}/2}\nonumber\\
&-&{\bm \Sigma}^{({\bf A})}_{{\bf k}+{\bf q}/2,{\bf k}-{\bf q}/2}(i\omega_m)]^{-1}, 
\end{eqnarray}
where ${\bm \Sigma}^{({\bf A})}$ denotes the electron self-energy. 
In the following we use the short-hand notation ${\bf k}_-\equiv{\bf k}-{\bf q}/2$ and ${\bf k}_+\equiv{\bf k}+{\bf q}/2$.

Taking the derivative of the energy $K$ in \eref{eq:K00app} as required by the definition of the orbital magnetization \eref{eq:morb2app} gives:
\begin{widetext}
\begin{multline}
{ M}^a_{orb}
=\frac{-i\epsilon^{abc}}{4N\beta}\sum_{{\bf k}\omega_m} {\rm Tr}\bigg\{\frac{\partial }{\partial {q}_b}\bigg(
\frac{\partial {\bm \lambda}^{E,({\bf A})}_{0,{\bf k}_-,{\bf k}_+}}{\partial {A}^c_{\bf q}}\bigg|_{{\bf A}={\bm 0}}{\bf G}_{{\bf k}_+,{\bf k}_-}
-{\bm \lambda}^{E}_{0,{\bf k}_-,{\bf k}_+}{\bf G}_{{\bf k}_+,{\bf k}_+}\frac{\partial{\bf G}^{({\bf A})-1}_{{\bf k}_+,{\bf k}_-}}{\partial {A}^c_{\bf q}}\bigg|_{{\bf A}={\bm 0}}{\bf G}_{{\bf k}_-,{\bf k}_-}\bigg)\bigg\}_{{\bf q}={\bm 0}}e^{i\omega_m0^+},
\label{eq:K1app}
\end{multline} 
\end{widetext}
where ${\bm \lambda}^{E,({\bf A})}_{0,{\bf k}_-,{\bf k}_+}\equiv [{\bf H}^{({\bf A})}_{0,{\bf k}_-,{\bf k}_+}+(i\omega_m-\mu){\bm \delta}_{{\bf q},{\bm 0}}]$ is the bare energy vertex. Its derivative with respect to the gauge potential gives the bare current vertex. The Green's function that multiplies this vertex must be evaluated at ${\bf A}={\bm 0}$ so it is diagonal in momentum space and ${\bf G}_{{\bf k}_+,{\bf k}_-}={\bm 0}$. In the last term of the equation, we have used the identity $(\partial {\bf G^{({\bf A})}}/\partial {A}^c_{\bf q})={\bf G}(-\partial {\bf G}^{({\bf A})-1}/\partial {A}^c_{\bf q}){\bf G}$. The derivative of the inverse of the Green's function with respect to the gauge potential is the dressed current vertex function which can be related to the bare current vertex using the Bethe-Salpeter equation. In the last term on the right-hand side of \eref{eq:K1app}, one can see that the dressed current vertex adds momentum $-{\bf q}$. The Green's functions on either side are evaluated at zero vector potential and hence are diagonal in momentum index. 

Performing the derivative with respect to $q_b$, keeping in mind that the first term in the above equation is identically zero, we find 
\begin{widetext}
\begin{eqnarray}
{ M}^a_{orb}=
\frac{-i\epsilon^{abc}}{4N\beta}\sum_{{\bf k}\omega_m} {\rm Tr}\bigg(
&-&{\bf \lambda}^{E}_{0,{\bf k}_-,{\bf k}_+}\bigg[\frac{\partial {\bf G}_{{\bf k}_+,{\bf k}_+}}{\partial {q}_b}\frac{\partial{\bf G}^{({\bf A})-1}_{{\bf k}_+,{\bf k}_-}}{\partial {A}^c_{\bf q}}\bigg|_{{\bf A}={\bm 0}}{\bf G}_{{\bf k}_-,{\bf k}_-}+{\bf G}_{{\bf k}_+,{\bf k}_+}\frac{\partial}{\partial {q}_b}\frac{\partial {\bf G}^{({\bf A})-1}_{{\bf k}_+,{\bf k}_-}}{\partial {A}^c_{\bf q}}\bigg|_{{\bf A}={\bm 0}}{\bf G}_{{\bf k}_-,{\bf k}_-}\nonumber \\
&+&{\bf G}_{{\bf k}_+,{\bf k}_+}\frac{\partial{\bf G}^{({\bf A})-1}_{{\bf k}_+,{\bf k}_-}}{\partial {A}^c_{\bf q}}\bigg|_{{\bf A}={\bm 0}}\frac{\partial {\bf G}_{{\bf k}_-,{\bf k}_-}}{\partial {q}_b}\bigg]\bigg)_{{\bf q}={\bm 0}}e^{i\omega_m0^+}.
\label{eq:K2app}
\end{eqnarray} 
\end{widetext}

In the limit ${\bf q}\rightarrow {\bm 0}$ and ${\bf A}_{\bf q}\rightarrow {\bm 0}$, we can replace $(\partial/\partial {A}^c_{\bf q})$ by $-(e/\hbar)\partial/\partial {k}_c$ and $(\partial/\partial{ q}_b)$ by $(\pm 1/2) (\partial/\partial {k}_b)$, depending on the momentum of the propagator line. After this replacement, we can see that the first and the last terms in the above equation are equal. Finally, using the identity $(\partial {\bf G}/\partial {q}_b)={\bf G}(-\partial {\bf G}^{-1}/\partial {q}_b){\bf G}$, we have the formula for the orbital magnetization, 

\begin{widetext}
\begin{eqnarray}
{ M}^a_{orb}&=&(\frac{e}{\hbar})(\frac{i\epsilon^{abc}}{4N\beta})\sum_{{\bf k}\omega_m} {\rm Tr}\bigg\{
[{\bf H}_0+(i\omega_m-\mu){\bm 1}]\bigg( {\bf G}\frac{\partial}{\partial {q}_b}\frac{\partial {\bf G}^{({\bf A})-1}_{{\bf k}_+,{\bf k}_-}}{\partial {A}^c_{\bf q}}\bigg|_{{\bf A}={\bm 0},{\bf q}={\bm 0}}{\bf G} 
+
{\bf G}(-\frac{\partial {\bf G}^{-1}}{\partial {k}_b}) {\bf G}(-\frac{\partial {\bf G}^{-1}}{\partial {k}_c}){\bf G} \bigg)\bigg\}e^{i\omega_m 0^+}.
\label{eq:K4app}
\end{eqnarray} 
\end{widetext}

The derivative of ${\bf G}^{({\bf A})-1}$ contains two terms, one is the derivative of ${\bf H}_0^{({\bf A})}$ and the other one is the derivative of the self-energy. The former term vanishes because there is no $\bf q$ dependence left once the derivative with respect to ${\bf A}$ is evaluated at ${\bf A}={\bm 0}$. We then define $(i\epsilon^{abc}e/2\hbar)(\partial^2 {\bf \Sigma}^{({\bf A})-1}/\partial {q}_b\partial {A}^c_{\bf q})$ by $({\partial {\bf \Sigma}^{({\bf B})-1}/\partial { B}^a})$.  Only the gauge invariant part of ${\bm \Sigma}^{\bf B}$ contributes to the derivative. The resulting formula can be used directly to obtain the orbital magnetization. However, it is also possible to rewrite the last term to obtain the form in the main text by recalling that the energy vertex ${\bf H}_0+(i\omega_m-\mu){\bm 1}$ can be written as ${\bf G}^{-1} + 2({\bf H}_0 -\mu{\bm 1})+{\bm \Sigma}$. In that case, the product between ${\bf G}^{-1}$ and the last term in the above equation leaves a term that is symmetric with respect to the current vertices and therefore vanishes due to the cross product.  

\subsection{Derivation II}
In this subsection we provide an alternative derivation for the orbital magnetization based on a generalization of the method introduced in Ref.~\onlinecite{PhysRevB.84.205137} to the case of interacting systems. As we will verify, in this method, it is the uniform magnetic field that appears explicitly, so that the formalism is manifestly gauge-invariant. In addition, despite the fact that the Hamiltonian is not translationally invariant, any measured quantity can be calculated in an explicitly translationally invariant manner. 

In position space, with ${\bf R}_{i\alpha}\equiv {\bf R}_{i}+{\bf r}_{\alpha}$ where ${\bf R}_{i}$ is the origin of $i$th unit cell and ${\bf r}_{\alpha}$ denotes the position of $\alpha$th ion within the unit cell, $K$ can be obtained from
\begin{multline}
K=\frac{1}{2N\beta}\sum_{{\bf R}_{i\alpha}{\bf R}_{j\alpha'}}\sum_{\omega_m}\\ {\rm Tr}\bigg([{\bf H}_{0,{\bf R}_{i\alpha}{\bf R}_{j\alpha'}}+(i\omega_m-\mu){\bm \delta}_{{\bf R}_{i\alpha}{\bf R}_{j\alpha'}}]{\bf G}_{{\bf R}_{j\alpha'}{\bf R}_{i\alpha}}\bigg)e^{i\omega_m0^+}. \label{eq:K0app}
\end{multline}    

In the presence of a small uniform magnetic field, the noninteracting Hamiltonian becomes 
${\bf H}_{0,{\bf R}_{i\alpha}{\bf R}_{j\alpha'}}\rightarrow({\bf H}_0+{\bf H}')_{{\bf R}_{i\alpha}{\bf R}_{j\alpha'}}\exp[(ie/\hbar)\int_{{\bf R}_{i\alpha}}^{{\bf R}_{j\alpha'}}{\bf A}({\bf r})\cdot d{\bf r}]$, where ${\bf H}'$ is some local perturbation that includes atomic diamagnetism, and $\bf A$ is the gauge potential. The line integral of the gauge potential follows a straight line from ${\bf R}_{i\alpha}$ to ${\bf R}_{j\alpha'}$. Since the correction to the Green’s function and the energy from ${\bf H}'$ is of order of $|{\bf B}|^2$, we ignore it from now on. Thus, in the presence of the field, the energy vertex in \eref{eq:K0app} is multipled by the Peierls phase, $\exp(i\phi_{{\bf R}_{i\alpha}{\bf R}_{j\alpha'}})\equiv \exp[(ie/\hbar)\int_{{\bf R}_{i\alpha}}^{{\bf R}_{j\alpha'}}{\bf A}({\bf r})\cdot d{\bf r}]$, and the Green's function should be evaluated in presence of the field. 

The linear response of the Green's function to the field can be obtained perturbatively as follow. The Green function satisfies the following equation,
\begin{multline}
\sum_{{\bf R}_{j\alpha'}}[(i\omega_m+\mu){\bm \delta}_{{\bf R}_{i\alpha}{\bf R}_{j\alpha'}}-{\bf H}_{0,{\bf R}_{i\alpha}{\bf R}_{j\alpha'}}]e^{i\phi_{{\bf R}_{i\alpha}{\bf R}_{j\alpha'}}}\\{\bf G}^{({\bf A})}_{{\bf R}_{j\alpha'}{\bf R}_{k\alpha''}}
-{\bm \Sigma}^{({\bf A})}_{{\bf R}_{i\alpha}{\bf R}_{j\alpha'}}{\bf G}^{({\bf A})}_{{\bf R}_{j\alpha'}{\bf R}_{k\alpha''}}=\delta_{{\bf R}_{i\alpha}{\bf R}_{k\alpha''}},
\label{eq:Gfnapp}
\end{multline}
where ${\bm \Sigma}^{\bf (A)}$ is electron self-energy and the superscript $({\bf A})$ indicates that the quantity must be calculated in the presence of the field, to distinguish from quantities ${\bf G}$ and ${\bm \Sigma}$ calculated at $\mathbf{B}=0$. Defining  $\tilde{{\bf G}}^{({\bf A})}_{{\bf R}_{i\alpha}{\bf R}_{j\alpha'}}$ and $\tilde{{\bf \Sigma}}^{({\bf A})}_{{\bf R}_{i\alpha}{\bf R}_{j\alpha'}}$ by ${\bf G}^{({\bf A})}_{{\bf R}_{i\alpha}{\bf R}_{j\alpha'}}=\tilde{{\bf G}}^{({\bf A})}_{{\bf R}_{i\alpha}{\bf R}_{j\alpha'}}e^{i\phi_{{\bf R}_{i\alpha}{\bf R}_{j\alpha'}}}$ and ${\bm \Sigma}^{({\bf A})}_{{\bf R}_{i\alpha}{\bf R}_{j\alpha'}}=\tilde{{\bm \Sigma}}^{({\bf A})}_{{\bf R}_{i\alpha}{\bf R}_{j\alpha'}}e^{i\phi_{{\bf R}_{i\alpha}{\bf R}_{j\alpha'}}}$,~\cite{JPSJ.74.2813} respectively, singles out the gauge independent quantities identified by a tilde. Indeed, we can rearrange the equation for $\tilde{{\bf G}}^{({\bf A})}_{{\bf R}_{i\alpha}{\bf R}_{j\alpha'}}$ and $\tilde{{\bf \Sigma}}^{({\bf A})}_{{\bf R}_{i\alpha}{\bf R}_{j\alpha'}}$ so that it is gauge invariant. It suffices to multiply both sides of \eref{eq:Gfnapp} by $e^{i\phi_{{\bf R}_{i\alpha},{\bf R}_{k\alpha''}}}$. The right-hand side remains unity while on the left the three phases combine together to give the magnetic flux threading through the triangle formed by the three points ${\bf R}_{i\alpha},{\bf R}_{j\alpha'}{\bf R}_{k\alpha''}$. Independently of the gauge then, we obtain
\begin{multline}
\sum_{{\bf R}_{j\alpha'}}[(i\omega_m+\mu){\bm \delta}_{{\bf R}_{i\alpha}{\bf R}_{j\alpha'}}-{\bf H}_{0,{\bf R}_{i\alpha}{\bf R}_{j\alpha'}}-\tilde{{\bm \Sigma}}^{({\bf A})}_{{\bf R}_{i\alpha}{\bf R}_{j\alpha'}}]\\\tilde{{\bf G}}^{({\bf A})}_{{\bf R}_{j\alpha'}{\bf R}_{k\alpha''}}
e^{(ie/2\hbar){\bf B}\cdot ({\bf R}_{j\alpha'}-{\bf R}_{i\alpha})\times({\bf R}_{k\alpha''}-{\bf R}_{j\alpha'})}=\delta_{{\bf R}_{i\alpha}{\bf R}_{k\alpha''}}.
\end{multline}

This last equation is gauge invariant and also translationally invariant.~\cite{PhysRevB.68.155114} It tells us, along with the theorem on the diagrammatic expansion of $\tilde{{\bf \Sigma}}^{({\bf A})}$~\cite{JPSJ.74.2813}, that $\tilde{{\bf G}}^{({\bf A})}$ and $\tilde{{\bf \Sigma}}^{({\bf A})}$ can depend only on $\mathbf{B}$, not on $\mathbf{A}$. This equation can thus be solved for $\tilde{{\bf G}}^{({\bf A})}$ to first order in $\bf B$ by expanding the self-energy and the exponential and then Fourier transforming. It is important to define the Fourier transform as $c_{{\bf R}_{i\alpha}}=(1/\sqrt{N})\sum_{\bf k}e^{i{\bf k}\cdot {\bf R}_{i\alpha}} c_{{\bf k}\alpha}$ so that the phase defined by $\bf k$ remains coherent even within a unit cell. This is consistent with the definition of the Peierl's phase. With this definition of the Fourier transform, we obtain,~\cite{note1}
\begin{eqnarray}
\tilde{{\bf G}}^{({\bf A})}_{\bf k} &=& {\bf G}_{\bf k}+B^a{\bf G}_{\bf k}
\left(\frac{\partial \tilde{{\bm \Sigma}}^{({\bm B})}_{\bf k}}{\partial B_a} \right)_{{\bf B}={\bm 0}}{\bf G}_{\bf k}\nonumber\\ 
&+&\frac{ie}{2\hbar}B^a\varepsilon^{abc}
{\bf G}_{\bf k}\left(\frac{\partial {\bf G}_{\bf k}^{-1}}{\partial k_b} \right)\left(\frac{\partial {\bf G}_{\bf k}}{\partial k_c} \right).
\label{eq:Gtildeapp}
\end{eqnarray}

In presence of the external field $\mathbf{B}$, $K$ must be calculated with the trace expression \eref{eq:K0app} but with the energy vertex multiplied by the Peierls phase $e^{i\phi_{{\bf R}_{i\alpha}{\bf R}_{j\alpha'}}}$. Combining that phase with ${\bf G}^{({\bf A})}_{{\bf R}_{i\alpha}{\bf R}_{j\alpha'}}$ shows that the gauge invariant quantity 
$\tilde{{\bf G}}^{({\bf A})}_{{\bf R}_{i\alpha}{\bf R}_{j\alpha'}}$ enters the observable $K$. Using \eref{eq:Gtildeapp} for $\tilde{{\bf G}}^{({\bf A})}(\mathbf{k},i\omega_m)$ to first order and the definition of the orbital magnetization, \eref{eq:morb1}, one obtains for the orbital magnetization of interacting systems presented in \eref{eq:morb3}.

\section{Non-interacting system}\label{App:AppendixB}

In the noninteracting case the orbital magnetization can be written as a summation over the occupied bands that decomposes the orbital magnetization into the orbital moments of the carriers plus a correction from the Berry curvature. Here we thus discuss the noninteracting limit of our equation for the orbital magnetization and show that it, in that case, it reduces to the modern theory of the magnetization.
 
Using the band representation of the Green's function, ${\bf g}^{(b)}_{{\bf k}}(i\omega_m)=[(i\omega_m-\mu){\bm 1}-{\bm \epsilon}_{{\bf k}}]^{-1}$ where ${\bm \epsilon}_{{\bf k}}$ is a diagonal matrix, one can rewrite the orbital magnetization as 
\begin{widetext}
\begin{eqnarray}
{M}^a_{orb}&=&(\frac{ie}{2\hbar })(\frac{\epsilon^{abc}}{N\beta})\sum_{{\bf k}\omega_m}{\rm Tr}\bigg(({\bf H}_0({\bf k})-\mu{\bm 1}){\bf g}^{(b)}_{{\bf k}}(i\omega_m)
(\frac{\partial {\bf H}_0({\bf k})
}{\partial {k}_b}){\bf g}^{(b)}_{{\bf k} }(i\omega_m)
(\frac{\partial {\bf H}_0({\bf k})
}{\partial{k}_c}){\bf g}^{(b)}_{{\bf k}}(i\omega_m)
\bigg)\nonumber \\ 
&=&(\frac{ie}{2\hbar })(\frac{\epsilon^{abc}}{N\beta})\sum_{{\bf k}} \sum_{n,m}
\sum_{\omega_m}\frac{(\epsilon_{n{\bf k}}-\mu)(\frac{\partial {\bf H}_0({\bf k})
}{\partial {k}_b})_{nm}(\frac{\partial {\bf H}_0({\bf k})
}{\partial{k}_c})_{mn}}{(i\omega_m+\mu-\epsilon_{n{\bf k}})(i\omega_m+\mu-\epsilon_{m{\bf k}})(i\omega_m+\mu-\epsilon_{n{\bf k}})}
\nonumber \\ 
&=&(\frac{ie}{2\hbar })(\frac{\epsilon^{abc}}{N\beta})\sum_{{\bf k}} \sum_{n,m}(\epsilon_{n{\bf k}}-\mu)\frac{\partial}{\partial \epsilon_{n{\bf k}}}
\sum_{\omega_m}\frac{(\frac{\partial {\bf H}_0({\bf k})
}{\partial {k}_b})_{nm}(\frac{\partial {\bf H}_0({\bf k})
}{\partial{k}_c})_{mn}}{(i\omega_m+\mu-\epsilon_{n{\bf k}})(i\omega_m+\mu-\epsilon_{m{\bf k}})}
\nonumber \\ 
&=&(\frac{ie}{2\hbar })(\frac{\epsilon^{abc}}{N})\sum_{{\bf k}} \sum_{n,m}(\epsilon_{n{\bf k}}-\mu)\frac{\partial}{\partial \epsilon_{n{\bf k}}}\bigg\{
\frac{(\frac{\partial {\bf H}_0({\bf k})
}{\partial {k}_b})_{nm}(\frac{\partial {\bf H}_0({\bf k})
}{\partial{k}_c})_{mn}}{(\epsilon_{n{\bf k}}-\epsilon_{m{\bf k}})}[n_F(\epsilon_{n{\bf k}}-\mu)-n_F(\epsilon_{m{\bf k}}-\mu)]\bigg\}
\nonumber \\ 
&=&(\frac{-ie}{2\hbar })(\frac{\epsilon^{abc}}{N})\sum_{{\bf k}} \sum_{n,m}(\epsilon_{n{\bf k}}-\mu)
\frac{(\frac{\partial {\bf H}_0({\bf k})
}{\partial { k}_b})_{nm}(\frac{\partial {\bf H}_0({\bf k})
}{\partial{k}_c})_{mn}}{(\epsilon_{n{\bf k}}-\epsilon_{m{\bf k}})^2}[n_F(\epsilon_{n{\bf k}}-\mu)-n_F(\epsilon_{m{\bf k}}-\mu)]\nonumber \\&+&
(\frac{ie}{2\hbar })(\frac{\epsilon^{abc}}{N})\sum_{{\bf k}} \sum_{n,m}(\epsilon_{n{\bf k}}-\mu)
\frac{(\frac{\partial {\bf H}_0({\bf k})
}{\partial {k}_b})_{nm}(\frac{\partial {\bf H}_0({\bf k})
}{\partial{k}_c})_{mn}}{(\epsilon_{n{\bf k}}-\epsilon_{m{\bf k}})}\big(\frac{\partial n_F(\epsilon_{n{\bf k}}-\mu)}{\partial \epsilon_{n{\bf k}}}\big).\label{eq:Morbnonint2app}
\end{eqnarray}
\end{widetext}
At zero temperature the term involving the derivative of the Fermi function vanishes because $(\partial n_F(\epsilon_{n{\bf k}}-\mu)/\partial \epsilon_{n{\bf k}})$ becomes $\delta(\epsilon_{n{\bf k}}-\mu)$. By interchanging the band indices $n$ and $m$ in the term coming from $n_F(\epsilon_{m{\bf k}}-\mu)$ and noting that the cross product is giving a minus sign as well, the orbital magnetization is given by 
\begin{widetext}
\begin{eqnarray}
{M}^a_{orb}&=& (\frac{-ie}{2\hbar })(\frac{\epsilon^{abc}}{N})\sum_{{\bf k}} \sum_{n,m}(\epsilon_{n{\bf k}}+\epsilon_{m{\bf k}}-2\mu)
\frac{(\frac{\partial {\bf H}_0({\bf k})
}{\partial { k}_b})_{nm}(\frac{\partial {\bf H}_0({\bf k})
}{\partial{k}_c})_{mn}}{(\epsilon_{n{\bf k}}-\epsilon_{m{\bf k}})^2}n_F(\epsilon_{n{\bf k}}-\mu)\nonumber \\&=&(\frac{-ie}{2\hbar })(\frac{\epsilon^{abc}}{N})\sum_{{\bf k}} \sum_{n}
\langle {\partial}_{{k}_b}u_{n{\bf k}}|[{\bf H}_0({\bf k})-\epsilon_{n{\bf k}}]|{\partial}_{{ k}_c}u_{n{\bf k}}\rangle
n_F(\epsilon_{n{\bf k}}-\mu)\nonumber \\ 
&+&(\frac{-ie}{2\hbar })(\frac{\epsilon^{abc}}{N})\sum_{{\bf k}} \sum_{n}2(\epsilon_{n{\bf k}}-\mu)
\langle {\partial}_{{ k}_b}u_{n{\bf k}}|{\partial}_{{ k}_c}u_{n{\bf k}}\rangle
n_F(\epsilon_{n{\bf k}}-\mu),\label{eq:Morbnonint3app}
\end{eqnarray}
\end{widetext}
where we have used $\langle u_{n{\bf k}}|{\bm\nabla}_{\bf k} {\bf H}_0({\bf k })|u_{m{\bf k}} \rangle = (\epsilon_{n{\bf k}}-\epsilon_{m{\bf k}})\langle {\bm\nabla}_{\bf k}u_{n{\bf k}}|u_{m{\bf k}} \rangle$. In the last identity, the first term is the orbital moments of carriers, while the second term is a correction from the Berry curvature. \cite{PhysRevLett.99.197202} The Berry curvature is given by ${\bm \Omega}_n({\bf k})=i{\bm\nabla}_{{\bf k}}\times\langle u_{n{\bf k}}|{\bm\nabla}_{{\bf k}}| u_{n{\bf k}}\rangle$, which is an intrinsic property of the band structure because it only depends on the wave function and can be interpreted as an effective magnetic field in momentum space.\cite{RevModPhys.82.1959} In a finite system, the Berry curvature correction gives the surface contribution to the orbital magnetization. 

Next we show that in the large lattice spacing limit the orbital moment contribution reduces to the conventional form. At the atomic site located at ${\bf R}_i$ in the crystal, we can define a set of Wannier orbitals $|w_{ni}\rangle = w_n({\bf r}-{\bf R}_i)$, so that the cell-periodic part of the (nonrelativistic) Bloch states are given by 
\begin{equation}
u_{n{\bf k}}({\bf r})
=\frac{1}{\sqrt{N}}\sum_i e^{-i{\bf k}\cdot({\bf r}-{\bf R}_i)} w_{ni}({\bf r}-{\bf R}_i).
\end{equation}
Substituting  the above equation in the orbital moment term, using the relation ${\bf v}=(-i/\hbar)[{\bf r},{\bf H}_0]$ and finally taking only the site diagonal matrix elements, $i = j$, of the Wannier functions  one obtains the following relation for the orbital moment \cite{PhysRevB.81.060409}
\begin{equation}
\mu_B\frac{1}{N}\sum_i\sum_n \langle w_{ni}|{\bf r}\times {\bf p}|w_{ni}\rangle n_F(\epsilon_{n{\bf k}}-\mu),
\end{equation}
where $\mu_B=(e\hbar /2m_e )$ is the Bohr magneton and where we have exploited the fact that the bulk states carry no net current, i.e., $\sum_n\langle w_{ni}|{\bf v}|w_{ni}\rangle=0$ and made the approximation ${\bf p}=m_e{\bf v}$. Clearly, in the limit of zero bandwidth (large lattice spacing) the Wannier functions
reduce to molecular (atomic) spin-orbitals, and this expression yields the standard usual free atom orbital angular momentum and the corresponding magnetic moment per atom.

Finally, we comment on the relation between the orbital magnetization and the Chern number. The Chern number is an integral of the Berry curvature over the first Brillouin zone.\cite{RevModPhys.82.1959} As can be seen from \eref{eq:Morbnonint3app} and from the fact that the states do not depend on chemical potential in the noninteracting system, the derivative of the orbital magnetization with respect to $\mu$ is proportional to the Chern number when we are in the insulating state.\cite{PhysRevB.74.024408}

\section{Phase transition in the Kane-Mele-Hubbard}\label{App:AppendixC}
In the presence of a Hubbard-type interaction, the KMH Hamiltonian has two phases: An interacting quantum spin Hall insulator and a trivial xy-AFM insulator ($\lambda_{SO}\ne 0$).  The easy-plane AFM order is the result of the interplay between the Hubbard interaction and the spin-orbit coupling. The nearest-neighbor hopping generates an isotropic AFM Heisenberg term $(4t^2/U)\sum_{\langle ij\rangle}{\bf S}_i\cdot {\bf S}_j$ in the strong coupling limit, while the next nearest-neighbor hopping due to spin-orbit coupling generates an anisotropic exchange term $(4\lambda_{SO}^2/U)\sum_{\langle \langle ij\rangle \rangle}(-S^x_iS^x_j-S^y_iS^y_j+S^z_iS^z_j)$.\cite{PhysRevB.82.075106} The $z$-term in the later exchange term favours antiparallel alignment of the spin on the next nearest neighbor sites; thus, it introduces a frustration to the nearest-neighbor AFM correlation expressed by the former exchange terms. On the other hand, the $xy$ term in the latter exchange term favors a ferromagnetic alignment, so no frustration is introduced. As a result, the exchange term coming from the spin-orbit coupling has a tendency to suppress the $z$-term of the AFM order. 

A transition from a quantum spin Hall state to a topologically trivial state can occur either via the closing of the bulk band gap, or via the breaking of time-reversal symmetry. In the KMH model, upon increasing the Hubbard repulsion, 
a transition from the quantum spin Hall phase to a gapped
Mott insulator with long-range magnetic order occurs at a critical
value $U_c/t$.~ \cite{PhysRevB.85.115132}  At the magnetic transition, the time-reversal symmetry underlying the topological protection of the quantum spin
Hall state is broken and a change of the topological
invariant from nontrivial to trivial occurs without closing
any gap.~\cite{0953-8984-25-14-143201}

\section{Kane-Mele model}\label{App:AppendixD}
\begin{figure}
\begin{center}
\includegraphics[width=\normwidth]{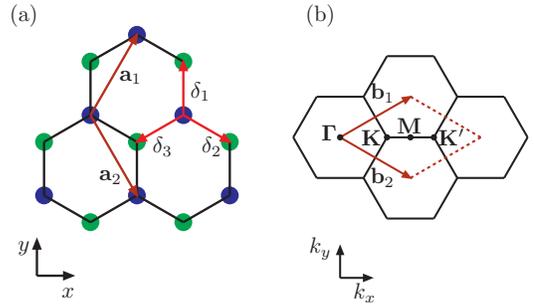}
\caption{(Color online) Panel (a): The honeycomb lattice with lattice constant $a$ consists
of two sublattices A, B and is spanned by the basis vectors 
${\bf a}_1 = a/2(\sqrt{3},3)$, ${\bf a}_2 = a/2(\sqrt{3},-3)$.
Nearest-neighbour lattice sites are connected by the vectors
${\bm \delta}_1 = a(0,1)$, ${\bm \delta}_2 = a/2(\sqrt{3},-1)$, and ${\bm \delta}_3 = a/2(-\sqrt{3},-1)$. Panel (b): The hexagonal first Brillouin zone contains the two nonequivalent Dirac
points ${\bf K} =(4\pi/3\sqrt{3}a)(1,0)$ and ${\bf K}^{\prime} =-(4\pi/3\sqrt{3}a)(1,0)$.}\label{honeycomb}
\end{center}
\end{figure}

In the absence of the  electron-electron interaction, \eref{eq:KMH} can be written in Fourier space in the form $\mathcal{H}_0^{\rm KM}=\sum_{\bf k}\Psi^{\dagger}_{\bf k}{\bf H}_0({\bf k})\Psi_{\bf k}$, with
\begin{equation}
\left(\begin{array}{cccc}
-\lambda+\lambda_{SO}\gamma_{\bf k} & -tg_{\bf k} & 0 & 0 \\
-tg^*_{\bf k} & \lambda-\lambda_{SO}\gamma_{\bf k} & 0 & 0 \\
0 & 0 & -\lambda-\lambda_{SO}\gamma_{\bf k} & -tg_{\bf k} \\
0 & 0 & -tg^*_{\bf k} & \lambda+\lambda_{SO}\gamma_{\bf k} 
\end{array}\right),\label{eq:Hk}
\end{equation}
where $\Psi^{\dagger}_{\bf k}\equiv(a^{\dagger}_{{\bf k}\uparrow},b^{\dagger}_{{\bf k}\uparrow},a^{\dagger}_{{\bf k}\downarrow},b^{\dagger}_{{\bf k}\downarrow})$. Here $a$ and $b$ operators refer to the two sublattices of the honeycomb lattice; $g_{\bf k}\equiv  \sum_i \exp(i{\bf k}.{\bm \delta}_i)$ is related to the nearest-neighbor hopping, with ${\bm \delta}_{i=1\cdots 3}$ denoting the three first-neighbor bond vectors; $\gamma({\bf k})=2\sum_i\sin({\bf k}.{\bf l}_i)$ where ${\bf l}_1={\bm \delta}_2-{\bm \delta}_3$, ${\bf l}_2={\bm \delta}_3-{\bm \delta}_1$ and ${\bf l}_3={\bm \delta}_1-{\bm \delta}_2$ (see \fref{honeycomb}).

The KM Hamiltonian, \eref{eq:Hk}, can be regarded as two decoupled models for the $\uparrow$ and $\downarrow$ spins, each equivalent to the spinless Haldane model, and described by $2\times2$ matrices. Although ${\bf H}_{\sigma}$ individually breaks time reversal symmetry, the complete Hamiltonian satisfies it.~\cite{0953-8984-25-14-143201} 
Furthermore, the centrosymmetric Hamiltonian  at half-filling ($\mu=0$) possesses the discrete particle-hole symmetry, $c^{\dagger}_{i\sigma}\rightarrow d_{i\sigma}=sc^{\dagger}_{i\sigma}, \;c_{i\sigma}\rightarrow d^{\dagger}_{i\sigma}=sc_{i\sigma}$ with $s=\pm 1$ depending on the sublatices. \cite{PhysRevB.84.205121} 

Any finite $\lambda_{SO}$ or $\lambda$ opens a bulk gap. The eigenvalues of KM Hamiltonian are
\begin{equation}
\epsilon_{\mp}({\bf k})=\mp\sqrt{t^2|g_{\bf k}|^2+(-\lambda+\lambda_{SO}\gamma_{\bf k})^2},
\end{equation}
so that the spectrum has two bands, each of which has a Kramers degeneracy between $\uparrow$ and $\downarrow$ spins. For $\lambda=0$, a bulk gap of size $\Delta=6\sqrt{3}\lambda_{SO}$ opens up at the Dirac points. For  $\lambda_{SO}/t>1/(3\sqrt{3})$ a minimal gap of size $\Delta=2t$ is instead found at the $ M=(2\pi/3a,0)$ point. For $\lambda_{SO}=0$ the charge gap is $\Delta=2\lambda$ at the Dirac points.~\cite{0953-8984-25-14-143201} 

The effective Dirac equation for states near the $\bf K$ and ${\bf K}'$ points is obtained from the following small $\bf q$ behavior of $g$ and $\gamma$: $g({\bf K}+{\bf q})\approx (3/2)a(q_x+iq_y)$, $g({\bf K}'+{\bf q})\approx (3/2)a(-q_x+iq_y)$, with $a$ the lattice spacing, and $\gamma({\bf K}+{\bf q})=-\gamma({\bf K}'+{\bf q})\approx 3\sqrt{3}$. The Hamiltonian can then be written as
\begin{align}
\boldsymbol{\mathcal{H}}({\bf q})&\equiv {\bf h}({\bf q}) \cdot {\bm \tau}\nonumber \\
&= \hbar v_F(s^v\tau_x q_x+\tau_y q_y)+(-\lambda + s s^v
3\sqrt{3}\lambda_{SO}) \tau_z,\label{eq:KMDirac}
\end{align}
acting on a two-component wavefunction with given spin that describes states on the $A$($B$) sublattice. In the above Hamiltonian, the valley index $s^v=\pm 1$ stands for states at the $\bf K$ ( ${\bf K}'$) points and $s=\pm 1$ represents spin direction. $\hbar v_F=(3/2)at$ is the Fermi velocity of the helical Dirac fermions. 

In the insulating phase of the KM Hamiltonian in the presence of an exchange field, only the Berry curvature correction contributes in the \emph{net} orbital magnetization. Equation (\ref{eq:KMDirac}) describes the low energy physics of the KM Hamiltonian in the insulating phase. Having the eigenstates, one can obtain an approximate analytical expression for the Berry curvature correction to the orbital magnetization integrand of each band around a given valley. The Berry curvature for each energy band is defined as ${\bm \Omega}_n({\bf q})=i{\bm\nabla}\times \langle u_n({\bf q})|{\bm\nabla}_{\bf q}|u_n({\bf q})\rangle$. Using the eigenstates $|u_{-}\rangle=[\exp(-i\phi)\sin(\theta /2), -\cos(\theta /2)]^T$ and  $|u_{+}\rangle=[\exp(-i\phi)\cos(\theta /2), \sin(\theta /2)]^T$, it can be shown that in two dimension the Berry curvature is given by \cite{RevModPhys.82.1959, RevModPhys.82.3045}
\begin{eqnarray}
{\Omega}_{\mp}^z({\bf q})&=&\pm i\frac{\sin\theta}{2}\left(\frac{\partial \theta}{\partial q_x} \frac{\partial \phi}{\partial q_y}-\frac{\partial \theta}{\partial q_y} \frac{\partial \phi}{\partial q_x}\right)\nonumber\\&=&
\pm \frac{i}{2}\frac{{\bf h}\cdot \partial_{q_x}{\bf h}\times \partial_{q_y}{\bf h}}{|{\bf h}|^3}.\label{eq:BC}
\end{eqnarray}
One verifies from \eref{eq:BC} that the Berry curvature is identically zero if $h_z=0$, i.e. for a centrosymmetric system without spin-orbit coupling.  For \eref{eq:KMDirac} with $(\partial h_x/\partial q_y)=(\partial h_y/\partial q_x)=(\partial h_z/\partial q_{x(y)})=0$, the above equation reduces to
\begin{equation}
{\Omega}_{\mp}^z({\bf q})=\pm i \frac{h_z}{2|{\bf h}|^3}\frac{\partial h_x}{\partial q_x}
\frac{\partial h_y}{\partial q_y}.\label{eq:BC2}
\end{equation}
This in turn gives the orbital magnetization integrand coming from the Berry curvature contribution as
\begin{eqnarray}
m_{orb}^{Berry}({\bf q})&=&(\frac{e}{4\hbar })\sum_{ss_v}[(\Delta^{s2}_v+\hbar^2 v_F^2q^2)^{1/2}+\mu] \nonumber \\
&\times& \frac{s^v\Delta^s_v\hbar^2 v_F^2}{[\Delta^{s2}_v+\hbar^2 v_F^2q^2]^{3/2}},\label{eq:analmapp}
\end{eqnarray}
where $\Delta^s_v=(-\lambda+ss^v3\sqrt{3}\lambda_{SO})$ is a valley and spin dependent gap.

An external exchange field adds the term $-sh{\bm 1}$ to the Hamiltonian, \eref{eq:KMDirac}. This perturbation does not change the Berry curvature as follows clearly from its definition \eref{eq:BC2}. However it linearly changes the energy vertex in the Berry curvature correction of the orbital magnetization. Thus the net orbital magnetization as a function of the exchange field is 
\begin{eqnarray}
M_{orb}(h)&=&h\sum_{\bf q}{\Omega}({\bf q})\nonumber\\&=&
h(\frac{e}{4\hbar })\sum_{\bf q}\sum_{ss_v}
\frac{s^v\Delta^s_v\hbar^2 v_F^2}{[\Delta^{s2}_v+\hbar^2 v_F^2q^2]^{3/2}},
\label{eq:morb}
\end{eqnarray} 
which is independent of Hamiltonian parameters.

%\bibliographystyle{prsty}
%\bibliography{manuscript}
%\end{thebibliography}


\end{document}